%% file: sn-arxiv_khoma.tex
\documentclass[sn-mathphys-num]{sn-jnl}% Math and Physical Sciences Numbered Reference Style 
\usepackage{graphicx}%
\usepackage{multirow}%
\usepackage{amsmath,amssymb,amsfonts}%
\usepackage{amsthm}%
\usepackage{mathrsfs}%
\usepackage[title]{appendix}%
\usepackage{xcolor}%
\usepackage{textcomp}%
\usepackage{manyfoot}%
\usepackage{booktabs}%

\begin{document}

\title[Elastic and charge transfer cross sections for low to ultralow $\rm{H}(1s) + \rm{H}^{+}$ collisions. Quantal and semiclassical calculations]{Elastic and charge transfer cross sections for low to ultralow $\rm{H}(1s)+\rm{H} ^{+}$ collisions. Quantal and semiclassical calculations}

\author*[1,2]{\fnm{Mykhaylo} \sur{Khoma}}\email{khomamv@nas.gov.ua}



\affil*[1]{\orgdiv{Max Planck Institute for the Physics of Complex Systems}, 
\orgaddress{\street{N\"othnitzer Str. 38}, \city{Dresden}, \postcode{01187}, \country{Germany}}}

\affil[2]{\orgdiv{Institute of Electron Physics}, \orgname{National Academy of Sciences of Ukraine}, 
\orgaddress{\street{Universytetska Str. 21}, \city{Uzhhorod}, \postcode{88017}, \country{Ukraine}}}

%\affil[3]{\orgdiv{Department}, \orgname{Organization}, \orgaddress{\street{Street}, \city{City}, \postcode{610101}, \state{State}, \country{Country}}}


\abstract{The elastic scattering and resonant charge transfer integral cross sections in $\rm{H}(1s) + \rm{H^+}$ collisions are computed for the center-of-mass energy range of $10^{-10}-10$ eV. Fully quantal and semiclassical approaches are utilized in these calculations. The reliability of the semiclassical approximation for very low collision energies is discussed. The results are compared with available data from the literature.}




\maketitle

\section{Introduction}\label{intro}

The elementary processes occurring during proton impact on atomic hydrogen are important for many applications such as astrophysical and fusion plasmas \cite{mnras-12,mnras-15,mnras-2012,Hodges-Geo91, Schultz-astro-05, Schultz-astro-08,Schultz-Stancil,Schultz-2021}, and have thus been a subject of active study for many decades  \cite{Hunter-77, Hunter-80, Krstic-APM-99, Krstic-PRA-60, Krstic-jpb-99, Krstic-PRA-70, Gao-PRA-2015, Kato-PRA}.

The fundamental quantum mechanical methods used to calculate the elastic and charge transfer cross sections for proton-hydrogen collisions in the low energy regime have been described in detail (see \cite{Hunter-77, Hunter-80, Krstic-APM-99,Krstic-PRA-60,Krstic-jpb-99} and references therein). These methods are based on the extraction of the phase shifts from the numerical propagation of the corresponding radial Schr\"odinger equation (or system of equations) to sufficiently large distances between the scattering particles. Having the phase shifts, the total or differential cross-sections can be computed in the standard manner \cite{Krstic-APM-99, Krstic-PRA-60, Krstic-jpb-99, Krstic-PRA-70}.

An alternative method which addresses the problem of the symmetric ion-atomic collisions is via the semiclassical computation of the scattering phases \cite{Kato-PRA, Dalgarno-WKB-53, Nakamura-ACP2007}. 
It is well known that the standard semiclassical approximations works reasonably well for intermediate and high collision energies. At low and very low collision energies, the so-called orbiting resonance phenomena may occur due to the possibility of formation and decay of the collision complex through trapping by the attractive polarisation force between the colliding particles \cite{Thylwe-and-Barany}. A more involved version (with three classical turning points) of the semiclassical theory exists to treat this case \cite{Connor-MolPhys,Connor-NATO}. Surprisingly, only a few works have directly applied this theory for computation of the cross sections in the presence of orbiting resonances \cite{Korsch-and-Thylwe, Thylwe-and-Barany}.

In this paper, we present detailed calculations of the phase shifts and the total cross sections 
for elastic scattering and charge transfer for symmetric $\rm{H}(1s) + \rm{H^+}$ collision
by means of fully quantal and semiclassical approaches. Numerical results are presented for the energy
range of $10^{-10} \le E_{\rm c.m.} \le 10$ eV. Here $E_{ {\rm c.m.} }$ is the collision energy in the center-of-mass frame. Finally, one of the goals of present work is to study the reliability of the semiclassical approach to ultra-low collision energies.

The article is organized as follows. In section 2, we briefly summarize 
the most relevant aspects of the theoretical approaches. The numerical details of the phase shift computations are also given in this section. In section 3, we present the results of the cross section calculations. The final discussion and concluding remarks are given in section 4. 

Throughout this paper we use atomic units (a.u.) for all quantities. For the sake of concreteness, 
we adopt the following energy conversion factor between a.u. and eV: eV/27.211386 \cite{CODATA-2014}.

\section{Theory}\label{theory}

We compute the cross sections for elastic scattering (EL) 
\begin{equation}\label{reaction-EL}
{\rm H}(1s) + {\rm H}^{+}  \to  {\rm H}(1s) + {\rm H}^{+}
\end{equation}
and for resonant charge transfer (CT) 
\begin{equation}\label{reaction-CT}
{\rm H}(1s) + {\rm H}^{+}  \to {\rm H}^{+}  + {\rm H} (1s) 
\end{equation}
for proton collision with atomic hydrogen using quantum mechanical and semiclassical methods.

For the energy range considered here, only two lowest electronic states of $\rm{H}_2^{+}$, 
the $1s\sigma_g$ {\it gerade} ($ g$) and $2p \sigma_u$  {\it ungerade} ($u$), are involved. 
Thus, the cross sections for reactions \eqref{reaction-EL} and \eqref{reaction-CT} may be obtained from solutions of uncoupled single-channel Schr\"odinger equations (see, e.g. \cite{Hodges-Geo91,Schultz-astro-05,Schultz-astro-08,Schultz-Stancil}) of the form
\begin{equation}\label{SE-radial-wf}
\left( \frac{ \rm{d}^2}{ {\rm{d}} R^2} + k^2 - 2\mu V^{(l)}_{g,u} (R) \right)
\psi^{(l)}_{g,u}(R) = 0,
\end{equation}
where $k^2={2\mu E_{ {\rm c.m.} }}$, $\mu$ is the system reduced mass, $l$ is the orbital angular momentum of the collision system, $R$ is the internuclear distance, $V^{(l)}_{g,u} (R)$ is the adiabatic potential for either the $1s \sigma_g$ or $2p\sigma_u$ electronic state,
and $\psi^{(l)}_{g,u}(R)$ is the radial wave function for the corresponding reaction channel. 
The adiabatic potential $V^{(l)}_{g,u} (R)$ reads
\begin{equation}\label{SE-Veff}
V^{(l)}_{g,u} (R) = E^{\rm BO}_{g,u}(R)  + \frac{l(l+1)}{2 \mu R^2},
\end{equation}
where $E^{\rm BO}_{g,u}(R)$ is the Born-Oppenheimer (BO) potential energy curve (PEC) of the corresponding  electronic state. For the purpose of the present study we calculated the BO PECs for $1s \sigma_g$ and $2p\sigma_u$ states of the $\rm{H}_2^+$ system to very high accuracy ($\sim$ 28-30 decimals) for a wide range of the distance $R$ ($0<R<10^{4}$ a.u.) by using a quadruple precision arithmetics and the algorithm described in \cite{mnras-12}.
We set the zero of energy at the asymptote of the BO PECs (i.e. $\lim_{R \to \infty} V^{(l)}_{g,u}(R) = 0$), so the asymptotic form of the radial wave functions 
$\psi^{(l)}_{g,u}(R)$ reads \cite{Le-Roy-06}
\begin{equation}\label{psi-rad-asymptote}
\psi^{(l)}_{g,u} (R) |_{R \to \infty} \simeq \sin( kR + \delta_{g,u}(l) - \pi l/2),
\end{equation}  
where $\delta_{g}(l)$ and $\delta_{u}(l)$ are the scattering phase shifts. 

\subsection{Semiclassical calculation} 

For not very low collision energies, the semiclassical (JWKB) approach can be used 
for computation of the scattering phase shifts \cite{Dalgarno-WKB-53,Krstic-jpb-99}. 
The 'classical' (single turning point $R_0$) JWKB approach provides the following expression for the phases $\delta_{g,u}(l)$
\begin{equation}\label{jwkb-delta}
\delta_{g,u}(l) =  
\int_{R_{0}}^{\infty} \Big( k_l(R) - k \Big) dR - k R_0 + \frac{\pi}{2} 
\left( l +\frac{1}{2} \right),
\end{equation}
where $k_l(R)$ is given by
\begin{equation}\label{jwkb-kl}
k_l(R) = \sqrt{ 2\mu  E_{\rm c.m.} - 2 \mu V_{l}^{\rm eff}(R)},
\end{equation}
and $V_{l}^{\rm eff}(R)$ is the effective potential including the Langer modification
\begin{equation}\label{jwkb-Veff}
V_{l}^{\rm eff}(R) = E^{\rm BO}_{g,u}(R)  + \frac{(l+1/2)^2 }{2 \mu R^2}.  
\end{equation}

%%%%%%%%%%%%%%%%%%%%%% three T-points %%%%%%%%%%%%%%%%%%%%%%%%%%%

As mentioned earlier in the Introduction, for collision systems with an attractive potential there can be a considerable range of $l$ values
for which the effective radial potential $V_{l}^{\rm eff}(R)$ has three classical turning points \cite{Connor-MolPhys}. We find this phenomena occurs 
for $1s \sigma_g$ state at collision energies $ E_{\rm c.m.} \lesssim 0.017$ a.u.

To determine the semiclassical phase shifts for this case we follow the algorithm and notations adopted in \cite{Connor-MolPhys,Connor-NATO}. Assume that $a_l$, $c_l$, and $e_l$ are the three classical turning points (in the order of increasing of $R$ value), and define the functions $\alpha_l$ and $\varepsilon_l$ as 
\begin{equation}\label{jwkb-alpha}
\alpha_l = \int_{a_l}^{c_l} k_l(R)dR,
\end{equation}
\begin{equation}\label{jwkb-eps}
\varepsilon_l = -\frac{1}{\pi} \int_{c_l}^{e_l} | k_l(R) |dR.
\end{equation}
Then, the sought phase shift $\delta(l)$ reads
\begin{equation}\label{jwkb-delta3p}
\delta(l) = \delta_{0}(l) - \frac{1}{2}\phi(\varepsilon_l) + \Theta (l),
\end{equation}
where $\delta_{0}(l)$ is determined by expression \eqref{jwkb-delta}, in which $R_0$ should be replaced by the outermost turning point $e_l$, the function $\phi(\varepsilon)$ is given by
\begin{equation}\label{jwkb-phi}
\phi(\varepsilon) = \varepsilon +  \arg \Gamma \left( \frac{1}{2} + i \varepsilon \right)
- \varepsilon \ln|\varepsilon|, 
\end{equation}
with the phase correction $\Theta (l)$ defined as follows:
\begin{equation}\label{jwkb-theta}
\Theta (l) = \arctan \left( \kappa(\varepsilon_l)
\tan [ \alpha_l - \frac{1}{2} \phi(\varepsilon_l) ]  \right),
\end{equation}
\begin{equation}
\kappa(\varepsilon)= 
\frac{ \sqrt{1+\exp(2\pi \varepsilon)} -1}{ \sqrt{1+\exp(2\pi \varepsilon)} +1}.
\end{equation}

The integration in \eqref{jwkb-delta} is carried out to sufficiently large $R_{ \rm max}$.
In all the semiclassical calculations presented here, we adopt $R_{\rm max} = 10^4$ a.u. This allows to achieve the convergence of $10^{-4}$ for all $\delta_{g,u}$. 

In Table \ref{table-rt} we show the coordinates of $a_l$, $c_l$, and $e_l$ for representative collision energies of $0.1$ eV and $0.2$ eV for all $l$ for which there exists three turning points. The results of the semiclassical calculations of the phase shifts and cross sections will be presented further on.  

%%%%%%%  table %%%%%%%%%%%%%5

\begin{table}[h]
\caption{Coordinates (in a.u.) of the turning points $a_l$, $c_l$, and $e_l$ for various $l$ (see text for details) and representative energies $E_{\rm{c.m.}}$ of $0.1$ eV and $0.2$ eV.}\label{table-rt}%
\begin{tabular}{@{}llll@{}}
\toprule
$l$    &  $a_l$    & $c_l$   & $e_l$ \\
\midrule
 $E_{\rm{c.m.}} = 0.1$ eV &&& \\
 27  &  1.968459  &  7.667607  &  9.664132\\
 28  &  2.040515  &  7.152738  & 10.346113\\
 29  &  2.117967  &  6.778401  & 10.890103\\
  30  &  2.201711  &  6.462717  & 11.376725\\
  31  &  2.292953  &  6.178178  & 11.831895\\
  32  &  2.393379  &  5.910682  & 12.267629\\
  33  &  2.505459  &  5.650765  & 12.690516\\
  34  &  2.633062  &  5.390169  & 13.104533\\
  35  &  2.782893  &  5.119235  & 13.512236\\
  36  &  2.968601  &  4.822214  & 13.915339\\
  37  &  3.228774  &  4.458966  & 14.315023\\
\midrule
$E_{\rm{c.m.}} = 0.2$ eV &&& \\      
  34 &   2.549141  &   6.667577  &  8.328068 \\
  35 &   2.676997  &   6.108765  &  8.924692 \\
  36 &   2.825851  &   5.681096 &   9.373407 \\
  37  &  3.007278  &  5.284696  &  9.762399 \\
  38  &  3.250047  &  4.864168  &  10.118181 \\
  39  &  3.715295  &  4.246132 &  10.452863 \\  
\botrule
\end{tabular}
%%\footnotetext{Source: This is an example of table footnote}
\end{table}

\subsection{Quantal calculation}

One of the most widely used methods for quantal calculation of the scattering phase shifts
consists of numerically integrating the radial equations \eqref{SE-radial-wf} and matching the obtained solutions to the asymptotic form \eqref{psi-rad-asymptote} (see, e.g., \cite{Hunter-77, Hunter-80,Krstic-APM-99,Thijssen,Gianozzi,Le-Roy-06}). In the present work, we use the approach described
in \cite{Thijssen, Gianozzi} to determine the phase shifts $\delta_{g,u}(l)$ at 1360 exponentially incremented energies between $10^{-10}$ eV and 10 eV. The energy points were chosen in order to resolve (as much as possible) the oscillating features of the integral cross sections. The standard Numerov's method was used to numerically integrate the radial equations \eqref{SE-radial-wf} from $R_{\rm min}$=0.001 a.u. to 
$R_{\rm max}$=1000 a.u. with a stepsize of  $R ~\sim~$0.001 a.u. In the lower energy regime, $E_{\rm c.m.} \leq 10^{-6}$ eV, the value of $R_{\rm max}$ was gradually increased from 5000 a.u. for $E_{\rm c.m.} = 10^{-6}$ eV up to $6 \times 10^4$ a.u. for $E_{\rm c.m.} = 10^{-10}$ eV. The choice of the largest orbital momentum $l_{\rm max}$ depends on the value of $E_{\rm c.m.}$ and was similar to those adopted in \cite{Schultz-astro-05}. Specifically, values of $l_{\rm max}$ vary from 9 at $10^{-10}$ eV, 28 at $10^{-8}$ eV, and up to 1200 for 10 eV. 
%
%table
\begin{table}[h]
\caption{Phase shifts for $E_{\rm{c.m.}}$ of $0.1$ eV and $10$ eV for representative values of $l$ from the present fully quantal (QUANT), present semiclassical (JWKB), and obtained by quantal calculations of  \cite{Hunter-80} (HK) and \cite{Krstic-jpb-99} (KS) where available.}\label{table-ph}
\begin{tabular*}{\textwidth}{@{\extracolsep\fill}lcccccccc}
\toprule%
& \multicolumn{2}{@{}c@{}}{QUANT} & \multicolumn{2}{@{}c@{}}{JWKB} & \multicolumn{2}{@{}c@{}}{HK} & \multicolumn{2}{@{}c@{}}{KS} \\\cmidrule{2-3}\cmidrule{4-5}\cmidrule{6-7}\cmidrule{8-9} %
  & g & u & g & u & g & u & g & u \\
\midrule
$ 0.1$ eV &&&&&&&& \\
$l=0$  & 1.221709 & 0.022883 & 1.22153 &  0.02799  & 1.234585 & 0.019202 & 1.21644 & 0.02638 \\
$l=5$  & 0.772645 & 0.624396 & 0.77276 &  0.62978  & 0.786002 & 0.620891 & 0.76734 & 0.62678 \\
$l=10$ & 3.303447 & 5.898259 & 3.30454 &  5.90441  & 3.317307 & 5.895139 & 3.29828 & 5.90031 \\
$l=20$ & 4.803677 & 5.473222 & 4.81175 &  5.48179  & 4.821474 & 5.471833 & 4.79954 & 5.47412 \\
$l=30$ & 0.572940 & 0.279244 & 0.57827 & 0.28236  & 0.574716  & 0.279115 & &  \\
$l=35$ & 0.279796 & 0.230648 & 0.28564 & 0.23139  & 0.280776  & 0.230459 & &  \\
$l=40$ & 0.173485 & 0.165080 & 0.17313 & 0.16517  & 0.175269  & 0.164588 & 0.17359 & 0.16505 \\
$l=60$ & 0.049876 & 0.049869 & 0.04985 & 0.04984  & 0.050337  & 0.050155 & 0.04990 & 0.04986 \\                         
$l=100$& 0.010815 & 0.010815 & 0.01080 & 0.01080  & 0.010871  & 0.010871 & 0.01082 & 0.01081 \\
\midrule
$10$ eV &&&&&&&& \\
$l= 0$& 3.020760 & 3.347830 & 3.018120 & 3.345851 & 3.023478 & 3.332009 & 3.01732 & 3.35621 \\
$l=10$& 2.237167 & 4.768027 & 2.235319 & 4.766228 & 2.239886 & 4.752823 & 2.23355 & 4.77620 \\
$l=20$& 2.243118 & 3.299371 & 2.242186 & 3.297947 & 2.245989 & 3.285574 & 2.23933 & 3.30703 \\
$l=40$& 0.127586 & 0.035456 & 0.127475 & 0.034722 & 0.131346 & 0.024905 & 0.12421 & 0.04172 \\
$l=100$& 0.236404 & 0.854549 & 0.236683 & 0.854526 & 0.241409 & 0.850499 & 0.23642 & 0.85710 \\
$l=200$& 0.502376 & 6.080151 & 0.502316 & 6.080164 & 0.502988 & 6.080018 & 0.50262 & 6.08038 \\
$l=300$& 0.055114 & 0.028372 & 0.055085 & 0.028352 & 0.064165 & 0.037268 & 0.05512 & 0.02835 \\
\botrule
\end{tabular*}
%\footnotetext{Note: This is an example of table footnote}
\end{table}

We find very good agreement in comparing our phase shifts with corresponding values tabulated in \cite{Hunter-77, Hunter-80}. The results are given in Table \ref{table-ph} for 0.1 and 10 eV 
for representative partial waves $l$. The integral cross sections for the reactions \eqref{reaction-EL} and \eqref{reaction-CT} are then computed from the phase shifts.

A note should be added concerning the definitions of elastic scattering and charge transfer for low-energy collisions involving identical nuclei as in the present case (a detailed discussion on this issue can be found in \cite{Hunter-77, Hunter-80, Krstic-APM-99, Krstic-jpb-99, Krstic-PRA-60, Krstic-PRA-70}).
In view of quantum indistinguishability of identical particles (QIP) the elastic scattering and charge transfer can only be differentiated by using a nuclear spin for labeling. Consequently, in the QIP approach the charge transfer is termed as spin exchange \cite{Hunter-77,Hunter-80,Krstic-APM-99,Krstic-PRA-60,Krstic-PRA-70,Krstic-jpb-99,Schultz-Stancil}. In this approach, the elastic scattering cross section contains the contributions both from the forward (elastic) and from the backward (charge transfer) scattering. With increase of collision energy the classical distinguishability of particles (CDP) becomes enabled, 
and so the elastic scattering and charge transfer processes can be separated out. 
Owing to the substantial interference of the direct and exchange scattering amplitude, 
the EL cross section in the QIP approach $\sigma^{\rm (i)}_{\rm EL}$ differs significantly from the EL cross section in the CDP approach $\sigma^{\rm (d)}_{\rm EL}$. Hence, two 
different sets of elastic scattering cross sections are produced in the present work for the symmetric collision system $\rm{H} + \rm{H}^{+}$  and compared to the results from  \cite{Schultz-Stancil,Krstic-PRA-70,Krstic-jpb-99,Kato-PRA}. In terms of the phase shifts $\delta_{g,u} (l)$ the integral cross sections $\sigma^{\rm (i)}_{\rm EL}$ and $\sigma^{\rm (d)}_{\rm EL}$ in the QIP and CDP approaches, respectively, are as follows (see, e.g., Eqs. (80) and (97) in \cite{Krstic-APM-99}):
\begin{equation}\label{sigma-elastic-inds}
\sigma^{\rm (i)}_{\rm EL} = \frac{\pi}{k^2} \Bigg\{  \sum_{l=0}^{ \rm even} (2l+1)
 \Big[ \sin^2 \delta_{g}  + 3 \sin^2 \delta_{u} \Big]
+ \sum_{l=1}^{{ \rm odd}} (2l+1)
 \Big[ \sin^2 \delta_{u} + 3 \sin^2 \delta_{g} \Big] \Bigg\},
\end{equation}
\begin{equation}\label{sigma-elastic-dst}
\sigma^{\rm (d)}_{\rm EL} = \frac{\pi}{k^2} \Bigg\{  \sum_{l}^{l_{\rm max}}(2l+1)
\Big[ \sin^2 \delta_{g}  +  \sin^2 \delta_{u} + 2 \sin \delta_{g} \sin \delta_{u} 
\cos( \delta_{g} - \delta_{u} )
\Big]  \Bigg\}.
\end{equation}
The charge transfer (CDP picture) and spin exchange (QIP picture) integral cross sections are identical and computed by means of the equation
\begin{equation}\label{sigma-cx-total}
\sigma_{\rm CT} = \frac{\pi}{k^2} \sum_{l=0}^{l_{ \rm max}} (2l+1) \sin^{2} 
\left( \delta_{g} - \delta_{u} \right).
\end{equation}
To simplify the notation we drop the argument $l$ of $\delta_{g,u}(l)$ in \eqref{sigma-elastic-inds}, \eqref{sigma-elastic-dst}, and \eqref{sigma-cx-total}.
From here forward we use the subscript $\rm CT$ for either charge transfer or spin exchange. In the high energy limit the EL cross section in the QIP picture tends to the sum of the EL and CT cross sections in the CDP picture, $\sigma^{\rm (i)}_{\rm EL}  = \sigma^{\rm (d)}_{\rm EL} + \sigma_{\rm CT}$ \cite{Krstic-PRA-60,Krstic-PRA-70,Krstic-jpb-99}.

\section{Results}

The very good agreement ($\sim 0.1\%$) of our calculations with the available results from \cite{Schultz-Stancil} (see the supplement materials to that publication) has been obtained,
except for the narrow range of $E_{\rm c.m.}$ from $2.2\times 10^{-4}$ to $2.8\times10^{-4}$ eV, where our results differ from those of \cite{Schultz-Stancil} by $2.5-8\%$ (see Table \ref{table-05}).

%--table------
\begin{table}[h]
\caption{The total cross sections in atomic units ($a_0^2=2.8003 \times 10^{-17} \rm{cm}^2 $) for charge transfer (CT) and for elastic scattering in the distinguishable (ELd) and indistinguishable (ELi) particles approaches for representative $E_{\rm c.m.}$ energies between $10^{-4}$ and $10^{-3}$ eV from the present work compared to the results from \cite{Schultz-Stancil}. The numbers in parentheses denotes the powers of ten to be multiplied.}\label{table-05}
\begin{tabular*}{\textwidth}{@{\extracolsep\fill}lcccccc}
\toprule%
%& \multicolumn{3}{@{}c@{}}{Element 1\footnotemark[1]} & \multicolumn{3}{@{}c@{}}{Element 2\footnotemark[2]} \\\cmidrule{2-4}\cmidrule{5-7}%
$E_{\rm{c.m.}}$(eV) & \text{CT}\footnotemark[1] & \text{CT}\footnotemark[2] & \text{ELd}\footnotemark[1] & \text{ELd}\footnotemark[2] & \text{ELi}\footnotemark[1] & \text{ELi} \footnotemark[2] \\
\midrule
 1.000($-$4) &  4048.3   &  4047 &  7839.9 & 7832 &10174.1 & 10179 \\
  1.096($-$4) &  3959.3   &  3958 &  8188.4 & 8176 &10273.1 & 10270 \\                               
  1.202($-$4) &  4055.2   &  4054 &  8600.5 & 8582 &10482.8 & 10470 \\                               
  1.349($-$4) &  4416.8   &  4415 &  8577.2 & 8560 &10578.7 & 10570  \\                               
  1.549($-$4) &  4544.0   &  4546 &  7178.9 & 7167 & 9829.8 & 9816  \\                               
  1.778($-$4) &  4111.9   &  4114 &  5846.6 & 5819 & 8807.8 & 8781  \\                               
  1.950($-$4) &  3765.4   &  3765 &  5445.3 & 5398 & 8328.0 & 8288  \\                               
  2.089($-$4) &  3562.5   &  3548 &  5445.0 & 5359 & 8143.7 & 8076 \\                               
  2.188($-$4) &  3518.5   &  3471 &  5662.9 & 5517 & 8174.7 & 8060 \\                               
  2.399($-$4) &  4427.4   &  4068 & 6182.9  & 6266 & 8781.1 & 8624 \\                               
  2.692($-$4) &  3563.8   &  3739 &  3467.2 & 3532 & 6874.9 & 6976  \\                                
  3.090($-$4) &  2846.2   &  2895 & 3403.4  & 3344 & 6396.9 & 6373  \\                               
  3.467($-$4) &  2577.7   &  2605 & 3451.3 & 3397  & 6273.4 & 6242  \\                                 
  4.571($-$4) &  2244.7  &  2256 & 3523.2  & 3487 & 6235.5 & 6210   \\                               
  5.248($-$4) &  2093.3  &  2100 &  3488.5 & 3462 & 6101.4 & 6081  \\                               
  6.026($-$4) &  1892.1  &  1895 &  3447.5 & 3429 & 5855.5 & 5840  \\                               
  6.918($-$4) &  1662.7  &  1663 &  3485.5 & 3473 & 5646.9 & 5635  \\                               
  8.128($-$4) &  1409.3  &  1408 &  3667.0 & 3659 & 5572.7 & 5563  \\                               
  9.772($-$4) &  1181.5  &  1178 & 3885.2  & 3885 & 5537.9 & 5533 \\
\botrule
\end{tabular*}
%\footnotetext{Note: This is an example of table footnote}
\footnotetext[1]{present calculation}
\footnotetext[2]{Ref. \cite{Schultz-Stancil}}
\end{table}

For $E_{\rm c.m.}=10^{-4}$ eV the present elastic total cross section $\sigma^{\rm (i)}_{\rm EL}$ is $10174.09 ~a_0^2$ compare to $10179.34 ~a_0^2$ obtained in \cite{Schultz-Stancil}, for $E_{\rm c.m.}=10^{-1}$ eV this cross section is $782.564 ~a_0^2$ compared to $783.0893 ~a_0^2$ from \cite{Schultz-Stancil}. 

\begin{figure}[b]
\includegraphics [width=0.9\textwidth,angle=0]{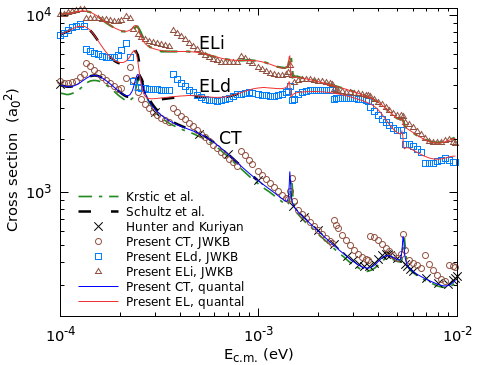}
\caption{\label{fig:1} Charge transfer (CT), elastic distinguishable (ELd), and elastic indistinguishable (ELi) integral cross sections from the present fully quantal (solid lines) and JWKB (symbols 
%$\bigcirc$, $\Circle$,
$\bigcirc$, $\square$, and $\triangle$) calculations 
for center-of-mass collision energies within the range of $10^{-4}$--$10^{-2}$ eV. 
Other theoretical results: $-\cdot-$, the $\sigma_{\rm{CT}}$ and $\sigma^{\rm (i)}_{\rm{EL}}$ calculations of Krsti{\'c} \textit{et al.} \cite{Krstic-PRA-70}; $-~-$, the $\sigma_{\rm{CT}}$ and $\sigma^{\rm (d)}_{\rm{EL}}$ calculations of Schultz \textit{et al.} \cite{Schultz-Stancil}; $\times$, the 
$\sigma_{\rm{CT}}$ calculations of Hunter and Kuriyan \cite{Hunter-77}
}
\end{figure}

\begin{figure}
\includegraphics [width=0.9\textwidth]{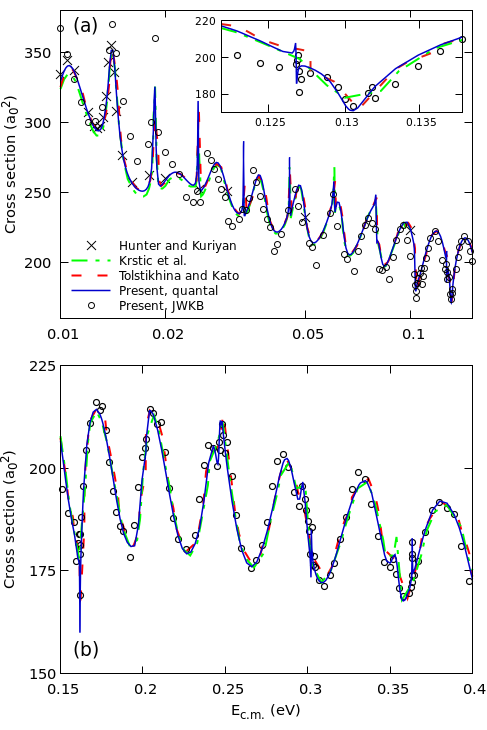}
\caption{\label{fig:2} 
(a) Charge transfer integral cross section from the present fully quantal (solid line) and JWKB (symbols 
%$\bigcirc$, $\Circle$,
$\bigcirc$) calculations within the range of $0.01$--$0.15$ eV. Other theoretical results: $\times$,  \cite{Hunter-77}; $-\cdot-$,  \cite{Krstic-PRA-70}; $-~-$,  \cite{Kato-PRA}. The inset shows a close-up of the sharp oscillation at $\sim$ 0.127 eV. (b) The same as in (a), except for the energy range of $0.15$--$0.4$ eV}
\end{figure}

\begin{figure}
\includegraphics [width=0.9\textwidth]{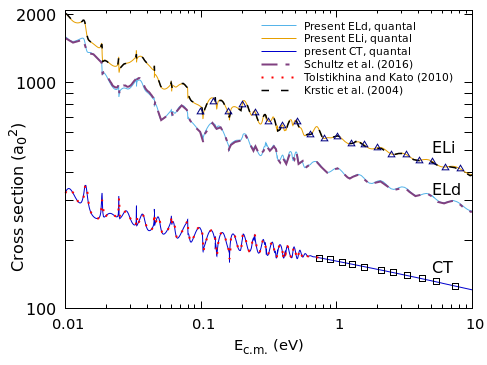}
\caption{\label{fig:4} Charge transfer (CT), elastic distinguishable (ELd), and elastic indistinguishable (ELi) integral cross sections for center-of-mass collision energies within the range of $0.01$--$10$ eV. Solid lines, the present quantal calculations; $-~-$, the quantal calculations from \cite{Schultz-Stancil}; $-\cdot-$, the quantal calculations from  \cite{Krstic-PRA-70}; 
$\cdots$, the quantal calculations from \cite{Kato-PRA}; $\bigtriangleup$ and $\square$, the JWKB calculations for ELi and CT from \cite{Krstic-jpb-99} 
}
\end{figure}

\begin{figure}[b]
\includegraphics [width=0.9\textwidth,angle=0]{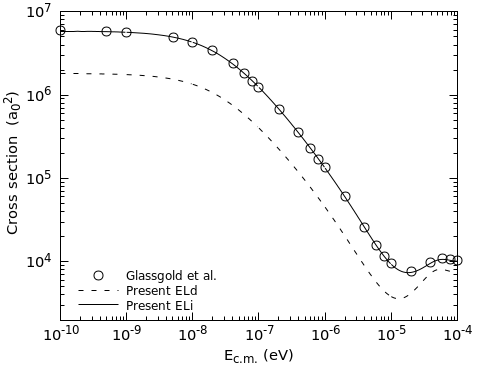}% Here is how to import EPS art
\caption{\label{fig:0} Integral elastic cross sections for the energy range of $10^{-10}$--$10^{-4}$ eV. Lines, the present calculations of $\sigma^{\rm (d)}_{\rm{EL}}$ and $\sigma^{\rm (i)}_{\rm{EL}}$; symbols $\bigcirc$, the $\sigma^{\rm (i)}_{\rm{EL}}$ calculations from \cite{Schultz-astro-05}
}
\end{figure}

Our calculated elastic and charge transfer integral cross sections over the center-of-mass energy range from $10^{-4}$ to $10^{-2}$ eV are displayed in Fig. \ref{fig:1}, where we have included other theoretical results from \cite{Krstic-PRA-70, Schultz-Stancil, Hunter-77}. Figure \ref{fig:2}(a,b) shows
the integral charge transfer cross section $\sigma_{\rm CT}$ from the present quantal and JWKB calculations for the energy range of $0.01-0.4$ eV. 
Notice the oscillatory patterns in the cross section (at $E_{\rm c.m.} \sim$ 0.127 and 0.363 eV) not reported in previous studies. The inset to Fig. \ref{fig:2}(a) shows a close-up of the sharp oscillation in the cross section at $E_{\rm c.m.} \sim$ 0.127 eV. The present results are compared with the theoretical data from 
 \cite{Hunter-77, Krstic-PRA-70, Kato-PRA, Schultz-Stancil}. Our quantal calculations of the elastic scattering cross sections $\sigma^{\rm (i)}_{\rm EL}$ and $\sigma^{\rm (d)}_{\rm EL}$ are shown in Fig. \ref{fig:4} over the energy range from $0.01$ to $10$ eV. Comparison was made with the quantal calculations from \cite{Schultz-Stancil, Krstic-PRA-70} and the JWKB calculations from  \cite{Krstic-jpb-99}. Finally, the present results of the elastic scattering cross sections at ultra-low collision energies
$10^{-10} \leq E_{\rm c.m.} \leq 10^{-5}$ eV are displayed in Fig. \ref{fig:0} 
along with the calculations from \cite{Schultz-astro-05}. Clearly, for 
such low collision energies, the distinguishable particles approximation fails completely and ELd results are given only for comparison purposes.
%

%%%%%%%%% figures %%%%%%%%%%%

\clearpage

\section{Discussion and concluding remarks}\label{sec12}

We have calculated the elastic and the resonant charge transfer cross sections in the collisions of proton with hydrogen atom $\rm{H}(1s)$. The calculations were performed over the range of the center-of-mass collision energies $10^{-10} - 10$ eV on a dense mesh of 1360 energy points. This allowed us to resolve all the known structures in the integral cross sections and to resolve the new ones at $E_{\rm c.m.} \sim 0.127$ and $0.363$ eV which were not reported in previous studies.
For the computation of the scattering phase shifts (and consequently the integral cross sections), we used both the fully quantal and the JWKB methods. The agreement of the results obtained by these two methods is excellent for the energies $E_{\rm c.m.} \gtrsim 0.47$ eV. For the energies $E_{\rm c.m.} \lesssim 0.47$ eV the standard (one-turning-point) JWKB method is not applicable to the considered collision system, and the three-turning-point version of the method must be used instead. We find that JWKB describes the cross sections quite well, both qualitatively and quantitatively, down to the collision energies of $\sim 0.01$ eV. 
For even lower collision energies, the accuracy of JWKB method (as expected) decreases significantly, but it is still possible to obtain a reasonable estimate for the trend of the cross sections down to energies of $0.001$ eV and even below. 

To conclude, we have shown that properly using the semiclassical method for cross section calculation allows us to significantly expand the range of its application to very low collision energies. 
The obtained semiclassical results agree well with the fully quantal calculations and also reproduce correctly the oscillatory features in the integral cross sections of the reactions under consideration.

\backmatter

%\bmhead{Supplementary information}

\bmhead{Acknowledgements}

Author acknowledge the Visitors Program Fellowship from the Max Planck Institute for the Physics of Complex Systems (MPI-PKS) as well very grateful to MPI-PKS for their warm hospitality during the stay in Dresden. The comments of Dr. M. Eiles on a draft version are greatly appreciated. The major part of the computations were performed on the computer cluster in the Institute of Electron Physics (Uzhhorod). I thanks Prof. E. Remeta for many encouraging discussions and Dr. O. Papp for support with computing facilities. This study was partially supported by the U.S. Office of Naval Research Global (Grant
N 62909-23-1-2088).

%\section*{Data availability statement}
%This manuscript has no associated data, or the data will not be deposited. Data
%sets generated during the current study are available from
%the corresponding author on reasonable request.

\clearpage

\clearpage

%%\bibliography{sn-bibliography}% common bib file
\input sn-article-bibtest.bbl

\end{document}